\documentclass[12pt]{article}
\input epsf.sty
\usepackage{mathrsfs}
\usepackage{amsmath}
\usepackage{mathrsfs}
\usepackage{amssymb}
\usepackage{color}
\usepackage{psfrag}
\usepackage{graphicx}
\usepackage{subfigure}
\usepackage{rotating}

\newcommand{\bea}{\begin{eqnarray}}
\newcommand{\eea}{\end{eqnarray}}
\newcommand{\be}{\begin{equation}}
\newcommand{\ee}{\end{equation}}
\newcommand{\vs}[1]{\vspace{#1 mm}}

\renewcommand{\a}{\alpha}
\renewcommand{\b}{\beta}

\renewcommand{\d}{\delta}

\newcommand{\dsl}{\pa \kern-0.5em /}

\newcommand{\half}{\frac{1}{2}}
\newcommand{\pa}{\partial}

\newcommand{\nn}{\nonumber\\}

\begin{document}
\topmargin 0pt
\oddsidemargin 0mm

\begin{flushright}



\end{flushright}

\vspace{2mm}

\begin{center}
{\bf Non-susy D3 brane and an interpolating solution between AdS$_5$\\ black hole, AdS$_5$
soliton and a `soft-wall' gravity solution} 

\vs{10}

{Shibaji Roy\footnote{E-mail: shibaji.roy@saha.ac.in}}

 \vspace{4mm}

{\em

 Saha Institute of Nuclear Physics,
 1/AF Bidhannagar, Calcutta-700 064, India\\}

\end{center}

\vs{10}

\begin{abstract}
It is known from the work in \cite{Lu:2007bu} of Lu et. al. that the non-supersymmetric 
charged D3-brane (with anisotropies in time as well as one of the spatial directions of D3-brane) 
of type IIB string theory
is characterized by five independent parameters. By fixing one of the parameters and zooming
into a particular region of space-time we construct a four parameter family of solution in
AdS$_5$, which interpolates between AdS$_5$ black hole and AdS$_5$ soliton (when one of spatial 
directions in the Poincare coordinates is compact) by continuously changing
the parameters (there is no need to take a double Wick rotation as is usual to go from one 
solution to the other) from one set of values to another. We consider two cases. In the 
first case the dilaton is constant
for this transition and there are only three independent parameters, whereas in the second case
the dilaton varies and there are four independent parameters. In the latter case, the
solution interpolates between AdS$_5$ black hole, AdS$_5$ soliton as well as the so-called 
`soft-wall' gravity solution of AdS/QCD model. We also compare our solution to the previously 
obtained Constable-Myers solution which is helpful in generalizing the solution for other D$p$
(for $p\neq 3$) branes.          
\end{abstract}

\newpage

\section{Introduction}

String theory in the low energy limit not only admits supersymmetric or BPS brane solutions 
\cite{Horowitz:1991cd,Duff:1993ye} but also
non-supersymmetric or non-susy brane solutions \cite{Zhou:1999nm,Brax:2000cf,Lu:2004ms}. Both these 
solutions are asymptotically flat. However,
unlike the BPS branes which are always charged and
are characterized by a single parameter, non-susy branes could be either charged or chargeless and
are characterized by more than one parameter. Here we will mainly be concerned with non-susy D-branes
carrying RR charges. Isotropic (in the world brane directions) non-susy D-branes\footnote{Like BPS branes
it has been shown recently in \cite{Nayek:2015tta} that gravity gets decoupled on non-susy branes as well. Therefore, it makes
sense to talk about gauge/gravity duality even for the non-susy branes where the gauge theory living on the brane
would be a non-susy YM theory like QCD.} are characterized by 
three independent parameters, but the number increases with the number of anisotropic directions \cite{Lu:2005jc}. So,
for example, if the number of anisotropic directions of a non-susy D$p$-brane is $q$ ($q\leq p$), then 
the number of independent parameters characterizing the solution would be $q+3$. We would like to remark 
that non-susy branes have a naked singularity and therefore are allowed to have more parameters without
violating Birkhoff's theorem. In \cite{Lu:2007bu}, we considered non-susy D$p$-brane solutions which are anisotropic
in the time direction as well as one of the spatial directions of the brane and therefore the solutions
contain five independent parameters. Apart from a non-trivial dilaton, the solutions contain an RR
$(8-p)$-form field strength. Because of the anisotropic directions, these solutions can also be interpreted
as a magnetically charged non-susy D$p$-brane intersecting with chargeless D0-brane and D1-brane. In \cite{Lu:2007bu}, we
have shown how these solutions (with D1 brane direction compact) interpolate between black D$p$-brane 
and Kaluza-Klein (KK) bubble of nothing (BON) when three of the five parameters change continuously from 
one set of 
values to another. Only at these two points in the parameter space, the solutions do not have naked 
singularity, otherwise all the solutions are singular. In \cite{Lu:2007bu}, we have interpreted this interpolation
as a transition triggered by the closed string tachyon condensation following \cite{Horowitz:2005vp}.  These solutions are string theory 
generalizations of the two parameter singular solutions obtained by Gross and Perry \cite{Gross:1983hb} in five dimensional KK
gravity which interpolates between KK black hole and KK BON \cite{Witten:1981gj}.

In this paper we consider a charged non-susy D3-brane with anisotropies in both time and one of the spatial 
directions of the brane of type IIB string theory. As we mentioned this solution is asymptotically flat and 
is characterized by five independent parameters. Now to construct asymptotically AdS$_5$ solution from here
we fix one of the five parameters of the solution and zoom into a particular region of space-time. The resulting
solution is a four-parameter asymptotically locally AdS$_5$ solution. This solution in general has non-trivial
dilaton. However by fixing one more parameter we can make the dilaton constant and the resulting three-parameter
solution interpolates between AdS$_5$ black hole and AdS$_5$ soliton (when one of the spatial directions in Poincare coordinates 
is compact) \cite{Horowitz:1998ha} by changing one of the parameters from
one specific value to another. There is no need to take double Wick rotation as is usual to go from black hole
to soliton solution. On the other hand, if we do not fix the parameter to make the dilaton constant, then the
four parameter solution interpolates between AdS$_5$ black hole, AdS$_5$ soliton and the `soft wall' gravity solution
of AdS/QCD model\footnote{The so-called `hard wall' gravity solution, where dilaton remains constant, was introduced in \cite{Polchinski:2001tt} 
to understand the high energy hard scattering behavior of QCD from string theory using gauge/string duality. However, this model
did not capture the linear Regge behavior of hadronic excitations as expected of a theory with linear confinement like QCD.
So, a `soft-wall' gravity solution with varying dilaton was introduced for this purpose in \cite{Karch:2006pv}. This model breaks the
conformal invariance and behaves much like QCD. Motivated by this many such solutions where dilaton was dynamically coupled to 
gravity were introduced and various QCD-like properties in them were studied \cite{Csaki:2006ji,Shock:2006gt,Kim:2007qk}.
Also see \cite{Nojiri:1999gf} for some earlier work on holographic QCD with varying dilaton.}. 
In this interpolation two of the four parameters change values continuously to take three sets
of values for the three different solutions. We remark that the three-parameter solution interpolating between
AdS$_5$ black hole and AdS$_5$ soliton has also been obtained before directly in five dimensional gravity with
negative cosmological constant \cite{Kiem:1998fy}. However, which ten dimensional brane solution it comes from is not clear. In this 
paper we clarify that the origin of this solution is actually the asymptotically flat, anisotropic non-susy D3 brane 
of type IIB string theory. The `soft wall' gravity solution \cite{Csaki:2006ji} which has been used as AdS/QCD model with running dilaton
is known as a solution of five dimensional gravity with a negative cosmological constant and a dilaton. The five
dimensional solution which interpolates between AdS$_5$ black hole and the `soft wall' solution is also known \cite{Kim:2007qk}. This is
also another three parameter solution different from the previous one which interpolates between AdS$_5$ black hole and AdS$_5$
soliton. Again their ten dimensional origin is not explicitly known. On the other hand our solution is a four parameter
solution (again with non-constant dilaton) and we show that it, in fact, interpolates 
between AdS$_5$ black hole, AdS$_5$ soliton and the `soft-wall' gravity solution. Similar ten dimensional string theory
solution with non-constant dilaton has been constructed by Constable and Myers \cite{Constable:1999ch}. We show how our solution maps to their
solution by a coordinate transformation and some redefinitions of parameters. So, implicitly our solution is known 
by Constable and Myers, and we here clarify how Constable-Myers solution can be regarded as asymptotically flat, 
anisotropic, non-susy D3-brane solution.  The gauge theory interpretations of these solutions are discussed in their 
paper, however, the interpolations of this solution is not clear there. The advantage of knowing the connection of
non-susy D3 brane with Constable-Myers solution is that, since the general non-susy D$p$ brane solutions are known \cite{Lu:2007bu}, 
they will give generalizations of Constable-Myers type solutions in other space-time dimensions.

This paper is organized as follows. In section 2, we briefly discuss the anisotropic non-susy D3 brane solution
characterized by five parameters. In section 3, we construct the four parameter asymptotically locally AdS$_5$ solution.
In subsection 3.1, we show how the solution interpolates bewteen AdS$_5$ black hole and AdS$_5$ soliton when the dilaton
is constant. In subsection 3.2, we consider the dilaton to be non-constant and show how the solution in this case
interpolates between AdS$_5$ black hole, AdS$_5$ soliton and the `soft wall' gravity solution. In section 4, we discuss
the relation between our solution and the Constable-Myers solution. Finally, we conclude in section 5. 

                      
\section{Anisotropic non-susy D3 brane solution}

The non-susy D$p$ brane solutions with anisotropies in time and one of the spatial directions of the branes of type II 
string theories are given in Eq.(4) of ref.\cite{Lu:2007bu}. Since the non-susy D$p$ branes have RR charge, the solutions can also be 
interpreted as charged D$p$ branes intersecting with chargeless non-susy D0 branes and D1 branes. For $p=3$, the solution
represents non-susy D3 branes with anisotropies in time as well as one of the spatial directions of the brane and will be
of our interest in this paper. The solution takes the form (putting $p=3$ in Eq.(4) of \cite{Lu:2007bu}),
\bea\label{nonsusyd3}
& & ds^2 = F^{\half} (H\tilde{H})^{\half}\left(\frac{H}{\tilde{H}}\right)^{\frac{3\d_1}{8}}\left(dr^2 + r^2 d\Omega_5^2\right)
+ F^{-\half}\left(\frac{H}{\tilde{H}}\right)^{\frac{3\d_1}{8}+ \d_0 + \half \d_2} (-dt^2)\nn
& & \qquad + F^{-\half} \left(\frac{H}{\tilde{H}}\right)^{-\frac{5\d_1}{8} + \d_0 - \frac{3\d_2}{2}} (dx^1)^2 + 
F^{-\half}\left(\frac{H}{\tilde{H}}\right)^{-\frac{5\d_1}{8} -\d_0 + \half \d_2}\sum_{i=2}^3 (dx^i)^2\nn
& & e^{2\phi} = \left(\frac{H}{\tilde{H}}\right)^{\frac{\d_1}{2} - 4\d_0 -2\d_2}, \qquad F_5 = (1+\ast) Q {\rm Vol}(\Omega_5)
\eea 
where the two harmonic functions $H(r)$ and $\tilde{H}(r)$ and the function $F(r)$ are defined as,
\bea\label{functions}
& & H(r) = 1 + \frac{\omega^4}{r^4},  \qquad \tilde{H}(r) = 1 - \frac{\omega^4}{r^4}\nn
& & F(r) = \left(\frac{H}{\tilde{H}}\right)^{\a} \cosh^2\theta - \left(\frac{\tilde{H}}{H}\right)^{\b} \sinh^2\theta
\eea
Here $\a$, $\b$, $\d_0$, $\d_1$, $\d_2$, $\theta$, $\omega$ and $Q$ are various integration constants and appear as the eight 
parameters of the solution. However, not all the parameters are independent. The consistency of the equations of motion restricts
some parameters by the following three contraints among them,
\bea\label{constraints}
& & \a-\b = -\frac{3}{2} \d_1\nn
& & \half \d_1^2 + \half \a\left(\a + \frac{3}{2}\d_1\right) + \half\left(2\d_2 + \d_0\right)\d_0 = \left(1-\d_2^2 - 2\d_0^2\right)\frac{5}{4}\nn
& & Q = 4 \omega^4 (\a+\b)\sinh2\theta
\eea  
Therefore, we can use these three relations to eliminate three parameters out of the eight just mentioned. The parameters $\a$, $\b$
can be expressed in terms of $\d_0$, $\d_1$ and $\d_2$ using the first two relations in \eqref{constraints}. 
So, the above solution
\eqref{nonsusyd3} contains five independent parameters, namely, $\d_0$, $\d_1$, $\d_2$, $\theta$ and $\omega$. Note that the metric
in \eqref{nonsusyd3} is given in the Einstein frame and it represents D3 brane with anisotropies in $t$ and $x^1$ directions as the
$dt^2$ and $(dx^1)^2$ terms have different coefficients from $(dx^2)^2, \, (dx^3)^2$ terms. $\phi$ is the dilaton and note that we have
suppressed the string coupling constant $g_s$ which is assumed to be small. $F_5$ is the RR self-dual 5-form field strength and $Q$ is
the charge of the non-susy D3 brane. 

Now for our purpose, we will use slightly different form of the solution than that given in \eqref{nonsusyd3}. For that we make a coordinate
change from $r$ to $\rho$ given by,
\be\label{coordchange}
r = \rho \left(\frac{1+\sqrt{f(\rho)}}{2}\right)^{\half}, \qquad {\rm with,} \quad f(\rho) = 1 - \frac{4\omega^4}{\rho^4} \equiv 
1 - \frac{\rho_0^4}{\rho^4}
\ee 
In this new coordinate we have,
\bea\label{newfunctions}
& & H = \frac{2}{1+\sqrt{f(\rho)}}, \qquad \tilde{H} = \frac{2\sqrt{f(\rho)}}{1 + \sqrt{f(\rho)}}, \qquad F = G(\rho) f^{-\frac{\a}{2}}\nn
& & dr = \frac{1}{\sqrt{f(\rho)}}\left(\frac{1 + \sqrt{f(\rho)}}{2}\right)^{\half} d\rho, \qquad dr^2 + r^2 d\Omega_5^2 = \frac{1+\sqrt{f}}{2}
\left(\frac{d\rho^2}{f} + \rho^2 d\Omega_5^2\right)
\eea
where,
\be\label{Ffunction}
G(\rho) = \cosh^2\theta - f^{\frac{\a+\b}{2}} \sinh^2\theta
\ee
Substituting \eqref{newfunctions} and \eqref{Ffunction} in \eqref{nonsusyd3}, we get the anisotropic non-susy D3 brane solution in
the following form,
\bea\label{nonsusyd3n}
& & ds^2 = G(\rho)^{\half} f(\rho)^{\frac{1}{4} - \frac{\a}{4} - \frac{3\d_1}{16}}\left(\frac{d\rho^2}{f(\rho)} + \rho^2 d\Omega_5^2\right) + G(\rho)^{-\half} 
f(\rho)^{\frac{\a}{4} - \frac{3\d_1}{16} - \frac{\d_0}{2} - \frac{\d_2}{4}} (-dt^2)\nn
& & \qquad + G(\rho)^{-\half} f(\rho)^{\frac{\a}{4} + \frac{5\d_1}{16} - \frac{\d_0}{2} + \frac{3\d_2}{4}} (dx^1)^2 + G(\rho)^{-\half} f(\rho)^{\frac{\a}{4} + \frac{5\d_1}{16} + 
\frac{\d_0}{2} - \frac{\d_2}{4}} \sum_{i=2}^3 (dx^i)^2\nn
& & e^{2\phi} = f(\rho)^{-\frac{\d_1}{4} + 2\d_0 + \d_2}, \qquad F_5 = (1+\ast) Q {\rm Vol}(\Omega_5)
\eea
The parameters of the solution satisfy the same relations as given before in \eqref{constraints}, where the last relation is now replaced by
\be\label{charge}
Q = \rho_0^4 (\a+\b)\sinh2\theta
\ee
This is the five parameter anisotropic non-susy D3 brane solution we will be concerned with in the following sections, where the five parameters, 
as mentioned before, are $\d_0$, $\d_1$, $\d_2$, $\rho_0$ and $\theta$. 

We would like to mention that isotropic BPS D3 brane solution can be recovered from \eqref{nonsusyd3n} if we take the double scaling limit
$\rho_0 \to 0$, $\theta \to \infty$ such that $\frac{\a+\b}{2}\rho_0^4 \sinh^2\theta = R^4$ = fixed. In that case $f(\rho) \to 1$,
$G(\rho) \to 1 + \frac{R^4}{\rho^4} = \bar{H}(\rho)$, the standard harmonic function of a BPS D3 brane and $Q$ in \eqref{charge} goes to
$4 R^4$. There is only a single parameter $R$ in the solution as expected.

\section{Asymptotically locally AdS$_5$ solution and interpolation}  

In this section we will construct a four parameter asymptotically locally AdS$_5$ solution from \eqref{nonsusyd3n}, by fixing one of the
five parameters characterizing the solution and zooming into a particular region of space-time. This four parameter solution has a non-constant
dilaton in general. However, the dilaton can be made constant if we fix another parameter leaving only a three parameter solution. This solution
will be shown to interpolate between AdS$_5$ black hole and AdS$_5$ soliton and will be discussed in subsection 3.1. When the dilaton is not constant
the four parameter solution will be shown to interpolate between AdS$_5$ black hole, AdS$_5$ soliton and a `soft wall' gravity solution 
and will be discussed
in subsection 3.2. When the interpolation includes AdS$_5$ soliton, the asymptotic geometry will include a circle (a compact direction along one of
the spatial directions in Poincare coordinates) with certain periodicity (discussed later in this section) to avoid conical singularity 
for the soliton solution. 

Now to obtain asymptotically locally AdS$_5$ solution\footnote{Another way to obtain asymptotically locally AdS$_5$ $\times$ S$^5$ solution, 
without putting $\a+\b=2$, is to do a scaling directly in the solution \eqref{nonsusyd3}. Actually if we take $\theta \to \infty$, the function
$F(r)$ given in \eqref{functions} takes the form $F(r) = \left[\left(\frac{H}{\tilde{H}}\right)^{\a} - \left(\frac{\tilde{H}}{H}\right)^{\b}\right]\cosh^2\theta$.
If we further scale $r \to \frac{r}{\cosh\theta}$ and $\omega \to \frac{\omega}{\cosh\theta}$, then the new metric $ds_{\rm n}^2 = (\cosh\theta)\,ds^2$
will be asymptotically ($r \to \infty$) locally AdS$_5$ $\times$ S$^5$. In obtaining this we also need to scale the charge as $Q \to \frac{Q}{\cosh^2\theta}$.
Similar scaling has been performed by Constable and Myers in \cite{Constable:1999ch}. The scaled solution can also be shown to interpolate 
between various AdS$_5$ solutions
we are discussing when the parameters take precisely the values we obtain in subsections 3.1 and 3.2. The difference between our asymptotically locally 
AdS$_5$ $\times$ S$^5$ solution and that of Constable-Myers is that 
in our case the radius of the five dimensional sphere is constant whereas for Constable-Myers it is not.}, 
we put $\a+\b=2$. Then the function $G(\rho)$ defined in \eqref{Ffunction} takes the form,
\be\label{Gfunction}
G(\rho) = \cosh^2\theta - f(\rho) \sinh^2\theta = 1 + \frac{\rho_0^4 \sinh^2\theta}{\rho^4} = 1 + \frac{R^4}{\rho^4}
\ee
If we now zoom into the region $\rho \sim \rho_0 \ll \rho_0 \sinh^{\half}\theta = R$, then $G(\rho) \approx \frac{R^4}{\rho^4}$. Therefore, the
solution in \eqref{nonsusyd3n} will take the form,
\bea\label{nonsusyd3n1}
& & ds^2 = \frac{R^2}{\rho^2}\frac{d\rho^2}{f(\rho)} + \frac{\rho^2}{R^2} f(\rho)^{\frac{1}{4} - \frac{3\d_1}{8} - \frac{\d_0}{2}-\frac{\d_2}{4}} (-dt^2) 
+ \frac{\rho^2}{R^2} f(\rho)^{\frac{1}{4} + \frac{\d_1}{8} - \frac{\d_0}{2} + \frac{3\d_2}{4}} (dx^1)^2\nn 
& & \qquad\qquad + \frac{\rho^2}{R^2} f(\rho)^{\frac{1}{4} + \frac{\d_1}{8} + 
\frac{\d_0}{2} - \frac{\d_2}{4}} \sum_{i=2}^3 (dx^i)^2 + R^2 d\Omega_5^2\nn
& & e^{2\phi} = f(\rho)^{-\frac{\d_1}{4} + 2\d_0 + \d_2}, \qquad F_5 = (1+\ast) Q {\rm Vol}(\Omega_5)
\eea
Note that since we have set $\a+\b=2$ and also from the first equation in \eqref{constraints} we have $\a-\b=-\frac{3}{2}\d_1$, we therefore
have $\a=1-\frac{3}{4}\d_1$ and we have used this in writing the metric in \eqref{nonsusyd3n1}. We now make a coordinate transformation
from $\rho$ to $z$ by, $\rho = \frac{R^2}{z}$, for which the function $f$ takes the form,
\be\label{f}
f = 1 - \frac{\rho_0^4}{\rho^4} = 1 - \frac{\rho_0^4 z^4}{R^8} \equiv 1 - \frac{z^4}{z_0^4} = f(z), \qquad {\rm where,}\quad z_0 = \frac{R^2}{\rho_0}
\ee
The solution \eqref{nonsusyd3n1} then takes the form,
\bea\label{nonsusyd3n2}
& & ds^2 = \frac{R^2}{z^2}\left[\frac{dz^2}{f(z)} + f(z)^{\frac{1}{4} - \frac{3\d_1}{8} - \frac{\d_0}{2}-\frac{\d_2}{4}} (-dt^2) 
+ f(z)^{\frac{1}{4} + \frac{\d_1}{8} - \frac{\d_0}{2} + \frac{3\d_2}{4}} (dx^1)^2\right.\nn 
& & \left. \qquad\qquad + f(z)^{\frac{1}{4} + \frac{\d_1}{8} + 
\frac{\d_0}{2} - \frac{\d_2}{4}} \sum_{i=2}^3 (dx^i)^2\right] + R^2 d\Omega_5^2\nn
& & e^{2\phi} = f(z)^{-\frac{\d_1}{4} + 2\d_0 + \d_2}, \qquad F_5 = (1+\ast) Q {\rm Vol}(\Omega_5)
\eea  
The parameter relations \eqref{constraints} now take the forms,
\bea\label{constraints1}
& & 7\d_1^2 + 96 \d_0^2 + 40 \d_2^2 + 32 \d_0\d_2 = 24\nn
& & Q = 4R^4
\eea
Note that we have already used the first relation in \eqref{constraints}. We therefore have four parameters in the solution \eqref{nonsusyd3n2}
namely, $\d_1$, $\d_2$, $R$ and $z_0$. Also note that by changing the coordinate from $\rho$ to $z$ we have changed the boundary from $\rho = \infty$
to $z=0$ and we recover the AdS$_5$ $\times$ S$^5$ metric from \eqref{nonsusyd3n2} when $z=0$. From the form of $f(z)$ given in \eqref{f} we notice
that there is a horizon at $z=z_0$, but the horizon is in general singular. The solution is physical only when $z<z_0$. Thus Eq.\eqref{nonsusyd3n2} 
represents the four parameter asymptotically locally AdS$_5$ solution. In the next two subsections we will show how this solution interpolates 
between various known AdS$_5$ solutions by fixing some of the parameters when dilaton is constant and when dilaton is not constant.

\subsection{Interpolation with constant dilaton} 

It is clear from the solution given in \eqref{nonsusyd3n2} that the dilaton will be constant if the parameters satisfy
\be\label{cond1}
\frac{\d_1}{4} = \d_2 + 2\d_0 \qquad \Rightarrow \quad \d_0 = \frac{\d_1}{8} - \frac{\d_2}{2}
\ee
Using this the five dimensional solution takes the form (see \eqref{nonsusyd3n2} without the S$^5$ factor),
\bea\label{ads5solution}
& & ds^2 = \frac{R^2}{z^2}\left[\frac{dz^2}{f(z)} + f(z)^{\frac{1}{4} - \frac{7\d_1}{16}} (-dt^2) 
+ f(z)^{\frac{1}{4} + \frac{\d_1}{16} + \d_2} (dx^1)^2 + f(z)^{\frac{1}{4} + \frac{3\d_1}{16} - \frac{\d_2}{2}} 
\sum_{i=2}^3 (dx^i)^2\right]\nn
& & e^{2\phi} = 1
\eea  
where $f(z)$ is as given in \eqref{f}. The parameter relation \eqref{constraints1} now takes the form,
\be\label{constraints2}
17\d_1^2 + 96 \d_2^2 -16 \d_1 \d_2 = 48
\ee
So, \eqref{ads5solution} is a three parameter solution with the independent parameters $\d_2$, $R$ and $z_0$. It is clear that the above 
solution will reduce to AdS$_5$ black hole solution if the parameters $\d_1$ and $\d_2$ satisfy,
\bea\label{bh}
& & \frac{1}{4} - \frac{7\d_1}{16} = 1\nn
& & \frac{1}{4} + \frac{\d_1}{16} + \d_2 = 0\nn
& & \frac{1}{4} + \frac{3\d_1}{16} - \frac{\d_2}{2} = 0
\eea
The solution of these equations is
\be\label{bhsoln}
\d_1 = -\frac{12}{7}, \qquad {\rm and} \qquad \d_2 = - \frac{1}{7}
\ee
One can easily check that with this solution the parameter relation \eqref{constraints2} is automatically satisfied. 
Note that at this point in parameter space the horizon becomes non-singular and the only independent parameters are $R$ and $z_0$ as it 
should be for AdS$_5$ black hole.

Similarly, we find that the above solution \eqref{ads5solution} reduces to AdS$_5$ soliton if the parameters $\d_1$ and $\d_2$ satisfy,
\bea\label{soliton}
& & \frac{1}{4} - \frac{7\d_1}{16} = 0\nn
& & \frac{1}{4} + \frac{\d_1}{16} + \d_2 = 1\nn
& & \frac{1}{4} + \frac{3\d_1}{16} - \frac{\d_2}{2} = 0
\eea  
The solution of these equations is
\be\label{solitonsoln}
\d_1 = \frac{4}{7}, \qquad {\rm and} \qquad \d_2 =  \frac{5}{7}
\ee
Again one can check that with this solution the parameter relation \eqref{constraints2} is satisfied. At this point in parameter
space the solution is completely regular. However, there is a conical singularity at $z=z_0$. The conical singularity is removed
by compactifying the $x^1$ coordinate with periodicity $\pi z_0$. Here again the solution has two independent parameters
$R$ and $z_0$.

Thus we have shown how the solution \eqref{ads5solution} interpolates between AdS$_5$ black hole and AdS$_5$ soliton. Note that in this
interpolation the dilaton remains constant and this is assured as long as the parameters $\d_1$ and $\d_2$ satisfy the relation
\eqref{constraints2}. It is, therefore, clear that in the parameter space $\d_1$ and $\d_2$ are changing continuously subject to the
constraint \eqref{constraints2} which is an equation of an ellipse. So, each point on that ellipse represents an asymptotically 
locally AdS$_5$ solution and only at the two points $(\d_1,\,\d_2) = (-\frac{12}{7},\,-\frac{1}{7})$ and $(\d_1,\,\d_2) = (\frac{4}{7},\,\frac{5}{7})$
we have AdS$_5$ black hole and AdS$_5$ soliton solution respectively. So, the AdS$_5$ black hole and the AdS$_5$ soliton are connected continuously
by other singular solutions lying on that ellipse, just mentioned, in the parameter space.

\subsection{Interpolation with non-constant dilaton}

When dilaton is not constant we take the five dimensional solution as that given in \eqref{nonsusyd3n2} without the S$^5$ factor
and let us write it here to show the interpolation,
\bea\label{nonsusyd3n3}
& & ds^2 = \frac{R^2}{z^2}\left[\frac{dz^2}{f(z)} + f(z)^{\frac{1}{4} - \frac{3\d_1}{8} - \frac{\d_0}{2}-\frac{\d_2}{4}} (-dt^2) 
+ f(z)^{\frac{1}{4} + \frac{\d_1}{8} - \frac{\d_0}{2} + \frac{3\d_2}{4}} (dx^1)^2\right.\nn 
& & \left. \qquad\qquad + f(z)^{\frac{1}{4} + \frac{\d_1}{8} + 
\frac{\d_0}{2} - \frac{\d_2}{4}} \sum_{i=2}^3 (dx^i)^2\right]\nn
& & e^{2\phi} = f(z)^{-\frac{\d_1}{4} + 2\d_0 + \d_2}
\eea  
Again $f(z)$ is as given in \eqref{f} and the parameters satisfy the relations given in \eqref{constraints1}. This is a four parameter
solution where the independent parameters are $\d_1$, $\d_2$, $R$ and $z_0$. It is clear from \eqref{nonsusyd3n3} that the above
solution will reduce to AdS$_5$ black hole if the parameters satisfy,
\bea\label{bh1}
& & \frac{1}{4} - \frac{3\d_1}{8} - \frac{\d_0}{2} - \frac{\d_2}{4} = 1\nn
& & \frac{1}{4} + \frac{\d_1}{8} - \frac{\d_0}{2} + \frac{3\d_2}{4} = 0\nn
& & \frac{1}{4} + \frac{\d_1}{8} + \frac{\d_0}{2} - \frac{\d_2}{4} = 0
\eea 
The solution of these equations is
\be\label{bh1solution}
\d_1 = -\frac{12}{7}, \qquad {\rm and} \qquad \d_0 = \d_2 = -\frac{1}{7}
\ee
One can check that with this solution the constraint equation \eqref{constraints1} is automatically satisfied. Also note that with
this solution the dilaton in \eqref{nonsusyd3n3} becomes constant as it should be. So, in this case two of the four independent parameters
get fixed and the AdS$_5$ black hole solution has two free parameters as expected.

Now to get AdS$_5$ soliton solution from \eqref{nonsusyd3n3} the parameters must satisfy,
\bea\label{soliton1}
& & \frac{1}{4} - \frac{3\d_1}{8} - \frac{\d_0}{2} - \frac{\d_2}{4} = 0\nn
& & \frac{1}{4} + \frac{\d_1}{8} - \frac{\d_0}{2} + \frac{3\d_2}{4} = 1\nn
& & \frac{1}{4} + \frac{\d_1}{8} + \frac{\d_0}{2} - \frac{\d_2}{4} = 0
\eea
The solution for this set of equations is
\be\label{soliton1solution}
\d_1 = \frac{4}{7}, \qquad \d_2 = \frac{5}{7} \qquad {\rm and} \qquad \d_0 = -\frac{2}{7}
\ee
Again one can check that this solution satisfies the constraint relation of the parameters given in \eqref{constraints1}. Dilaton in this
case also can be seen to become constant with this solution. Soliton solution is regular, as we mentioned before, if $x^1$ is made compact
with periodicity $\pi z_0$. Here also, the solution is characterized by two free parameters. Thus we find that the interpolation
between AdS$_5$ black hole and AdS$_5$ soliton can occur even if the dilaton varies, however, at these two particular points in the parameter space
the dilaton becomes constant. 

As the dilaton is not constant in this case we find that the solution \eqref{nonsusyd3n3} can also interpolate to another solution the
so-called `soft wall' gravity solution of some AdS/QCD model \cite{Csaki:2006ji}. To obtain this solution the parameters must satisfy,
\bea\label{softwall}
& & \frac{1}{4} - \frac{3\d_1}{8} - \frac{\d_0}{2} - \frac{\d_2}{4} = \frac{1}{4}\nn
& & \frac{1}{4} + \frac{\d_1}{8} - \frac{\d_0}{2} + \frac{3\d_2}{4} = \frac{1}{4}\nn
& & \frac{1}{4} + \frac{\d_1}{8} + \frac{\d_0}{2} - \frac{\d_2}{4} = \frac{1}{4}
\eea
From \eqref{softwall} we get,
\be\label{softwallsoln}
\d_0=-\half \d_1 = \d_2
\ee
So, the equations \eqref{softwall} do not completely fix the parameters. However, when we substitute \eqref{softwallsoln} to the constraint
equation \eqref{constraints1} we get,
\be\label{softwallsoln1}
\d_1 = \pm \frac{2\sqrt{6}}{7}, \qquad {\rm and} \qquad \d_0 = \d_2 = \mp \frac{\sqrt{6}}{7}
\ee
The solution \eqref{nonsusyd3n3} therefore takes the form,
\bea\label{nonsusyd3n4}
& & ds^2 = \frac{R^2}{z^2}\left[\frac{dz^2}{f(z)} + f(z)^{\frac{1}{4}}\left(-dt^2 + \sum_{i=1}^3 (dx^i)^2\right)\right]\nn 
& & e^{2\phi} = f(z)^{\sqrt{\frac{3}{2}}}
\eea 
Note that in writing the dilaton, we have taken only the lower sign of the solution \eqref{softwallsoln1} so that near $z=z_0$, the dilaton
remains small. Also we notice from the metric in \eqref{nonsusyd3n4} that this solution now has restored the full (3+1) dimensional Poincare 
invariance of a 3-brane which
is required for it to be a candidate of QCD gravity model. Since here the dilaton is non-constant leading to a running coupling constant of the
boundary theory, this solution is called a `soft wall' gravity solution \cite{Csaki:2006ji} as opposed to the `hard wall' solution 
\cite{Polchinski:2001tt} where the dilaton is constant.  
However, to cast this solution to a more familiar form \cite{Csaki:2006ji} of the standard
`soft wall' gravity solution of AdS/QCD, we need to go to another coordinate defined by
\be\label{anothercoord}
\hat{z} = z \left(\frac{1+\sqrt{f(z)}}{2}\right)^{-\half}, \qquad {\rm where}, \quad f(z) = 1 - \frac{\rho_0^4 z^4}{R^8} = 1 - \frac{z^4}{z_0^4}
\ee
It is clear from the above relation that $\hat{z}$ is actually related to our original radial coordinate $r$ (see Eq.\eqref{nonsusyd3}) by the
relation $\hat{z} = \frac{R^2}{r}$. Then we have,
\bea\label{newh}
& & H(r) = 1 + \frac{\omega^4}{r^4} = 1 + \frac{\omega^4 \hat{z}^4}{R^8} = 1 + \frac{\hat{z}^4}{\hat{z}_c^4} = H(\hat{z}) = \frac{2}{1+\sqrt{f(z)}}\nn 
& & \tilde{H}(r) = 1 - \frac{\omega^4}{r^4} = 1 - \frac{\omega^4 \hat{z}^4}{R^8} = 1 - \frac{\hat{z}^4}{\hat{z}_c^4} = \tilde{H}(\hat{z}) = 
\frac{2\sqrt{f(z)}}{1+\sqrt{f(z)}}
\eea
where we have defined $\hat{z}_c = \frac{R^2}{\omega}$. So, from \eqref{anothercoord} we have $\hat{z}_c = \sqrt{2}z_0$. Using \eqref{anothercoord}
and \eqref{newh} we can rewrite the `soft wall' gravity solution given in \eqref{nonsusyd3n4} as,
\bea\label{nonsusyd3n5}
& & ds^2 = \frac{R^2}{\hat{z}^2}\left[d\hat{z}^2 + \left(H(\hat{z})\tilde{H}(\hat{z})\right)^{\half}\left(-dt^2 + \sum_{i=1}^3 (dx^i)^2\right)\right]\nn
& & \qquad = \frac{R^2}{\hat{z}^2}\left[d\hat{z}^2 + \sqrt{1-\frac{\hat{z}^8}{\hat{z}_c^8}}\left(-dt^2 + \sum_{i=1}^3 (dx^i)^2\right)\right]\nn 
& & e^{2\phi} = \left(\frac{1 + \frac{\hat{z}^4}{\hat{z}_c^4}}{1 - \frac{\hat{z}^4}{\hat{z}_c^4}}\right)^{-2\sqrt{\frac{3}{2}}}
\eea 
This is precisely the `soft wall' gravity solution of AdS/QCD model obtained by Csaki and Reece in \cite{Csaki:2006ji} (see Eqs.(3.10) and (3.11) in their 
paper\footnote{Since the dilaton in this model is non-trivial and the metric components have power corrections to pure AdS$_5$ solution, this
has been attributed to the effects of gluon condensates and has been argued to provide a natural IR cut-off 
for confinement in the boundary theory \cite{Csaki:2006ji}. The gluon condensate has been calculated there and is shown to be related to $\hat{z}_c$, by the 
relation $\langle {\rm Tr}G^2\rangle
= 4\sqrt{3}\sqrt{\frac{R^3}{8\pi G_N}}\frac{1}{\hat{z}_c^4} = \frac{8}{\pi \hat{z}_c^4}\sqrt{3(N^2-1)}$.}, where dilaton has opposite sign and does
not remain small near $\hat{z}=\hat{z}_c$).
In terms of $\hat{z}$ coordinate the general solution \eqref{nonsusyd3n3} can also be written as\footnote{The following solution with $\d_0=\d_2$
matches precisely with the solution obtained by Kim et. al. in \cite{Kim:2007qk}. Note that in this case it is again a three parameter solution with the
parameter relation given by $7\d_1^2 + 168 \d_0^2 = 24$. When $\d_1=-\frac{2\sqrt{6}}{7}$ and $\d_0=\frac{\sqrt{6}}{7}$ it becomes a `soft wall'
gravity solution and when $\d_1=-\frac{12}{7}$ and $\d_0=-\frac{1}{7}$ it becomes AdS$_5$ black hole solution. So, for $\d_0=\d_2$ the solution
interpolates between `soft wall' gravity solution and AdS$_5$ black hole. In this case the boundary theory has been interpreted as describing 
gluon condensate at finite temperature \cite{Kim:2007qk}. This latter solution in slightly different coordinate system has been
shown \cite{Dey:2014yra} to interpolate between AdS$_5$ in the UV and a hyperscaling violating Lifshitz space-time in the IR.}, 
\bea\label{nonsusyd3n6}
& & ds^2 = \frac{R^2}{\hat{z}^2}\left[d\hat{z}^2 + \left(H(\hat{z})\tilde{H}(\hat{z})\right)^{\half}
\left(\frac{H(\hat{z})}{\tilde{H}(\hat{z})}\right)^{-\frac{\d_1}{4} - \d_0 +\frac{\d_2}{2}}\left\{\left(\frac{H(\hat{z})}
{\tilde{H}(\hat{z})}\right)^{\d_1+2\d_0}(-dt^2)\right.\right.\nn
& & \qquad\qquad \left.\left. +\left(\frac{H(\hat{z})}{\tilde{H}(\hat{z})}\right)^{2\d_0-2\d_2}(dx^1)^2 + \sum_{i=2}^3 (dx^i)^2\right\}\right]\nn 
& & e^{2\phi} = \left(\frac{H(\hat{z})}{\tilde{H}(\hat{z})}\right)^{\frac{\d_1}{2} - 4\d_0 - 2\d_2}
\eea 
One can easily check that for $\d_1 = -\frac{2\sqrt{6}}{7}$ and $\d_0=\d_2 = \frac{\sqrt{6}}{7}$, the above solution reduces exactly to the `soft wall'
gravity solution \eqref{nonsusyd3n5}. On the other hand for $\d_1=-\frac{12}{7}$ and $\d_0=\d_2=-\frac{1}{7}$, the solution \eqref{nonsusyd3n6} reduces
to 
\bea\label{bhnew}
& & ds^2 = \frac{R^2}{\hat{z}^2}\left[d\hat{z}^2 + H(\hat{z})\left(-
\left(\frac{\tilde{H}(\hat{z})}{H(\hat{z})}\right)^2 dt^2 + \sum_{i=1}^3 (dx^i)^2\right)\right]\nn 
& & e^{2\phi} = 1
\eea
This is the AdS$_5$ black hole solution used in \cite{Kim:2007qk}. Also for $\d_1=\frac{4}{7}$, $\d_0 = -\frac{2}{7}$ and $\d_2 = \frac{5}{7}$, the solution 
\eqref{nonsusyd3n6} reduces to  
\bea\label{solitonnew}
& & ds^2 = \frac{R^2}{\hat{z}^2}\left[d\hat{z}^2 + H(\hat{z})\left(- dt^2 + \left(\frac{\tilde{H}(\hat{z})}{H(\hat{z})}\right)^2 (dx^1)^2 + 
\sum_{i=2}^3 (dx^i)^2\right)\right]\nn 
& & e^{2\phi} = 1
\eea
This is the AdS$_5$ soliton solution in the new coordinate.

We have thus shown how the solution \eqref{nonsusyd3n3} interpolates between AdS$_5$ black hole, AdS$_5$ soliton and `soft wall' gravity solution of AdS/QCD
model when the dilaton is non-constant by continuously changing the values of two of the four parameters characterizing the solution subject to
the constraint \eqref{constraints1}. The constraint actually represents an ellipsoid and each point on the ellipsoid corresponds to a solution given
by \eqref{nonsusyd3n3}. The AdS$_5$ black hole, AdS$_5$ soliton and `soft wall' gravity solutions are just three points on the ellipsoid and they are connected
continuously by infinite number of solutions corresponding to the infinite possible values of the parameters.  

\section{Anisotropic non-susy D3 brane and Constable-\\
Myers solution}

In this section we will show how the Constable-Myers (CM) asymptotically flat solution given in Eqs.(5.5) -- (5.7) in \cite{Constable:1999ch} 
gets exactly mapped to 
the anisotropic non-susy D3 brane solution \eqref{nonsusyd3} we consider in section 1. We take the solution in Eqs.(5.5) -- (5.7) because this is 
more general than that in Eq.(2.1). The solution (2.1) is isotropic and is a scaled version of solution (5.5) -- (5.7). Moreover, since we were
considering the interpolation between AdS$_5$ black hole and AdS$_5$ soliton, our solution is isotropic in $x^2$ and $x^3$ directions. So, we will also take
CM solution isotropic in $y \equiv x^2$ and $z \equiv x^3$ directions by putting $\a_2=\a_3$. Also, not to confuse with the functions and the parameters
of our solution we will use a `bar' to denote CM solution. The Einstein frame metric and the dilaton for the CM solution have the forms,
\bea\label{CM}
& & ds^2 = \bar{F}(\bar{r})^{-\half}\left(-\bar{f}(\bar{r})^{\bar{\d}-\bar{\a}_1-2\bar{\a}_2} dt^2 + \bar{f}(\bar{r})^{\bar{\a}_1} (dx^1)^2 + \bar{f}(\bar{r})^{\bar{\a}_2} 
\sum_{i=2}^3 (dx^i)^2\right)\nn
& & \qquad + \bar{F}(\bar{r})^{\half} \bar{f}(\bar{r})^{\frac{2-\bar{\d}}{4}} \left[\frac{d\bar{r}^2}{\left(1+\frac{\omega^4}{\bar{r}^4}\right)^{\frac{5}{2}}} +
\frac{\bar{r}^2 d\Omega_5^2}{\left(1+\frac{\omega^4}{\bar{r}^4}\right)^{\half}}\right]\nn
& & e^{2\phi} = \bar{f}(\bar{r})^{\bar{\Delta}}
\eea     
where,
\be\label{Ff}
\bar{F}(\bar{r}) = \left(\bar{f}(\bar{r})^{\bar{\d}} - 1\right) \bar{\b}^2 + 1 \qquad {\rm and} \qquad \bar{f}(\bar{r}) = 1 + \frac{2\omega^4}{\bar{r}^4}
\ee
The parameters in the solution satisfy the relation,
\be\label{paramrelation}
\bar{\Delta}^2 + \frac{5}{2} \bar{\d}^2 - 4\bar{\d}(\bar{\a}_1 + 2\bar{\a}_2) + 4(\bar{\a}_1^2 + 3\bar{\a}_2^2 + 2\bar{\a}_1 \bar{\a}_2) = 10
\ee
Note that the parameter $\omega$ is the same in both the solutions. The radial coordinate in the two solutions are related as
\be\label{barr}
\bar{r}^4 = r^4 - \omega^4
\ee
In terms of $\bar{r}$ coordinate the functions $H(r)$ and $\tilde{H}(r)$ in our solution \eqref{nonsusyd3} take the forms,
\bea\label{htildeh}
& & H(r) = 1 + \frac{\omega^4}{r^4} = \frac{1+\frac{2\omega^4}{\bar{r}^4}}{1+\frac{\omega^4}{\bar{r}^4}}\nn
& & \tilde{H}(r) = 1 - \frac{\omega^4}{r^4} = \frac{1}{1+\frac{\omega^4}{\bar{r}^4}}
\eea 
We therefore have 
\be\label{htildeh1}
\frac{H(r)}{\tilde{H}(r)} = 1 + \frac{2\omega^4}{\bar{r}^4} = \bar{f}(\bar{r})
\ee
and so, the function $F(r)$ in our solution takes the form,
\bea\label{Four}
& & F(r) =  \left(\frac{H(r)}{\tilde{H}(r)}\right)^{\a} \cosh^2\theta - \left(\frac{\tilde{H}(r)}{H(r)}\right)^{\b} \sinh^2\theta\nn
& & \qquad = \left[\left(\bar{f}(\bar{r})^{\a+\b} -1 \right)\cosh^2\theta + 1\right]\bar{f}(\bar{r})^{-\b} = \bar{F}(\bar{r}) \bar{f}(\bar{r})^{-\b}
\eea
where we have identified $\a+\b = \bar{\d}$ and $\cosh^2\theta = \bar{\b}^2$. Further, substituting
\be\label{subs}
(H(r)\tilde{H}(r))^{\half} = \frac{\left(1 + \frac{2\omega^4}{\bar{r}^4}\right)^{\half}}{\left(1 + \frac{\omega^4}{\bar{r}^4}\right)}, 
\quad r^2 = \bar{r}^2 \left(1 + \frac{\omega^4}{\bar{r}^4}\right)^{\half}, \quad {\rm and} \quad dr^2 = 
\frac{d\bar{r}^2}{\left(1 + \frac{\omega^4}{\bar{r}^4}\right)^{\frac{3}{2}}}
\ee
we can write the solution \eqref{nonsusyd3} as follows,  
\bea\label{oursoln}
& & ds^2 = \bar{F}^{-\half}\left(-\bar{f}^{\frac{\a}{2}+\frac{9\d_1}{8}+\d_0+\frac{\d_2}{2}} dt^2 + \bar{f}^{\frac{\a}{2}+\frac{\d_1}{8}+\d_0-\frac{3\d_2}{2}} (dx^1)^2 + 
\bar{f}^{\frac{\a}{2}+\frac{\d_1}{8}-\d_0+\frac{\d_2}{2}} \sum_{i=2}^3 (dx^i)^2\right)\nn
& & \qquad + \bar{F}^{\half} \bar{f}^{-\frac{\a}{2}-\frac{3\d_1}{8}+\frac{1}{2}} \left[\frac{d\bar{r}^2}{\left(1+\frac{\omega^4}{\bar{r}^4}\right)^{\frac{5}{2}}} +
\frac{\bar{r}^2 d\Omega_5^2}{\left(1+\frac{\omega^4}{\bar{r}^4}\right)^{\half}}\right]\nn
& & e^{2\phi} = \bar{f}^{-\frac{\d_1}{2}-4\d_0-2\d_2}
\eea
where we have used the first relation in \eqref{constraints}.  
Comparing this solution \eqref{oursoln} with the CM solution given in \eqref{CM} we identify the parameters as,
\bea\label{paramid}
& & \bar{\a}_1 = \frac{\a}{2} + \frac{\d_1}{8} - \frac{3\d_2}{2} + \d_0\nn
& & \bar{\a}_2 = \frac{\a}{2} + \frac{\d_1}{8} + \frac{\d_2}{2} - \d_0\nn 
& & \bar{\d} = 2\a + \frac{3\d_1}{2}\nn
& & \bar{\Delta} = \frac{\d_1}{2} - 4 \d_0 - 2\d_2
\eea
Using these identifications one can easily check that the parameter relation \eqref{paramrelation} reduces to
\bea\label{paramrelation1}
& & \bar{\Delta}^2 + \frac{5}{2} \bar{\d}^2 - 4\bar{\d}(\bar{\a}_1 + 2\bar{\a}_2) + 4(\bar{\a}_1^2 + 3\bar{\a}_2^2 + 2\bar{\a}_1 \bar{\a}_2) = 10\nn
& & \Rightarrow \half \d_1^2 + \half \a(\a+\frac{3}{2}\d_1) + \half(2\d_2+\d_0)\d_0 = (1-\d_2^2-2\d_0^2)\frac{5}{4} 
\eea 
where the second line is precisely the parameter relation we have for anisotropic non-susy D3 brane solution given in the second equation in 
\eqref{constraints}. This, therefore, gives an exact mapping of anisotropic non-susy D3 brane solution to the CM solution. The 5-form field 
strength of CM solution is proportional to $(\bar{\b}^2-1)$, however, for anisotropic non-susy D3 brane it is proportional to 
$\sinh\theta\cosh\theta = \bar{\b}\sqrt{\bar{\b}^2-1}$ and the difference may be due to a typo. The solution (2.1) of Constable and Myers
\cite{Constable:1999ch} is, as we mentioned, a scaled solution of the general solution \eqref{CM}. One can check that when $\bar{\delta} = \a + \b =2$,
this solution reduces to the `soft wall' gravity solution given in \eqref{nonsusyd3n4}.

\section{Conclusion}

To conclude, in this paper we have constructed an asymptotically locally AdS$_5$ solution which interpolates between AdS$_5$ black hole
and AdS$_5$ soliton when the dilaton is constant and interpolates between AdS$_5$ black hole, AdS$_5$ soliton and `soft wall' gravity solution
of AdS/QCD model when dilaton is not constant. We obtained this solution from asymptotically flat, charged non-susy D3 brane solution with 
anisotropies in time and one of the spatial directions of the brane of type type IIB string theory. The asymptotically locally AdS$_5$ solution
is constructed as a throat limit of this solution (unlike many other solutions in the literature \cite{Nojiri:1999gf,Csaki:2006ji,Kim:2007qk,Kiem:1998fy}
which were obtained directly in AdS$_5$ space). Our solution has some similarities with the 
Constable-Myers (CM) solution \cite{Constable:1999ch}. We have shown how our asymptotically flat anisotropic non-susy D3 brane solution 
gets exactly mapped to the similar
solution constructed by CM. However, the asymptotically locally AdS$_5$ solution of CM is different from our solution in the sense that in
CM case the radius of the transverse 5-sphere is not constant, whereas, in our solution it is constant. CM solution is also a four parameter
solution (as in our case) because the parameter $\theta$ is scaled away whereas our solution is four parameter because we set $\a+\b=2$. The scaled
solution of CM as we have pointed out can also be shown to interpolate between AdS$_5$ black hole, AdS$_5$ soliton and `soft wall' gravity solution 
as in our solution at precisely those parameter values we described in section 3.     

In the process of showing the interpolation, we have clarified the relation of the various apparently unrelated AdS$_5$ solutions known in the 
literature \cite{Csaki:2006ji,Kim:2007qk,Kiem:1998fy,Constable:1999ch} and have seen that their ten dimensional origin is anisotropic non-susy 
D3 brane solution of type IIB string theory. The gauge theory
interpretations of these solutions have been discussed at length in many of these papers. To understand similar interpretations of our interpolating
solution, we note first of all that since there are no quarks, it will describe a pure glue Yang-Mills theory. In the first case, when the dilaton
is constant, the interpolation between AdS$_5$ soliton to AdS$_5$ black hole, which is a Hawking-Page like transition in gravity theory, is a 
confinement/deconfinement transition in the boundary theory. In the transition process the dilaton remains constant and the two states are
connected by infinite number of singular solutions. In \cite{Horowitz:2005vp}, this transition has been argued to be caused by the closed string tachyon
condensation and the singular solutions indicate our inability to describe this process (which is a stringy process) by purely classical
descriptions \cite{Lu:2007bu}. Similar discussion also applies for the case when dilaton is not constant. However, some new phenomenon occurs here in between.
In this case the interpolation occurs through a series of singular solutions where dilaton varies. From the form of the dilaton it is clear that
asymptotically this is a normalizable mode and therefore gives rise to the vaccum condensate $\langle{\rm Tr}G^2\rangle \neq 0$, which is the 
gluon condensate.
When $\d_1=-\frac{2\sqrt{6}}{7}$ and $\d_0=\d_2=\frac{\sqrt{6}}{7}$, we get `soft wall' gravity solution and then the solution describes gluon
condensate at zero temperature \cite{Csaki:2006ji}. However, for the general case, with only restriction that $\d_0=\d_2$ 
(otherwise arbitrary), the resulting
solution has been argued \cite{Kim:2007qk} to describe gluon condensate at finite temperature. When all the parameters are arbitrary, 
it can be checked that the
general solution has gluon condensate $\langle{\rm Tr}G^2\rangle \sim (\frac{\d_1}{2} - 4\d_0 - 2\d_2)$. As argued in \cite{Kim:2007qk}, 
the general solution
also has a temperature (despite the singular nature of the solution) which is related to the parameters as $T^4 \sim (\frac{\d_1}{2}+\d_0)$.                 
Also since the general solution \eqref{nonsusyd3n3} contains a compact direction for it to interpolate between AdS$_5$ soliton and AdS$_5$ black hole, 
the boundary theory
must contain Casimir energy \cite{Horowitz:1998ha} which can be shown to be related to the parameters as $\langle T_{tt}\rangle \sim (-\d_1+4\d_0-6\d_2)$.  
This, therefore, gives the physical meaning to the parameters of supergravity solution in terms of the physical quantities in the boundary theory.

The standard way to introduce quarks in our system is to look at the dynamics of the fundamental strings ending on D3 branes or introduce D7 branes
\cite{Karch:2002sh} in the D3 brane background and look at its dynamics. The resulting system can describe strongly coupled quark-gluon plasma. 
Since the non-susy D3 brane
we have is anisotropic\footnote{A regular gravity solution of type IIB string theory of anisotropic N=4 super YM plasma has been obtained in 
\cite{Mateos:2011ix}.}, 
so the system will describe anisotropic quark-gluon plasma which is thought to be generated in heavy ion collision at some
time scales earlier than when the isotropy sets in. One can study various transport properties of such system and see
how anisotropy affects them. Also, as we mentioned before, the advantage of the identification of our solution with Constable-Myers solution is that
since the general non-susy anisotropic D$p$ brane solutions are known \cite{Lu:2007bu}, we will have generalizations of CM solution in 
other space-time dimensions. 
This will help us to understand the nature of QCD-like theories in other dimensions using holography. We hope to come back on some of 
these issues in future. 
  
\vspace{.2cm}

\section*{Acknowledgements}
We would like to thank Somdeb Chakraborty, Jian-Xin Lu and Sang-Jin Sin for useful discussions.

\end{document}